\pdfoutput=1

\documentclass{aa}

\usepackage{graphicx}
\usepackage[varg]{txfonts}
\usepackage{amsmath}
\usepackage{verbatim}
\usepackage{latexsym}
\usepackage{multirow}
\usepackage{fixltx2e}
\usepackage{color}

\begin{document}

\title{Measuring solar active region inflows with local correlation tracking of granulation}

\author{B. L\"optien\inst{1,2}
\and A.~C. Birch\inst{2}
\and T.~L. Duvall Jr.\inst{2}
\and L. Gizon\inst{1,2}
\and B. Proxauf\inst{2}
\and J. Schou\inst{2}}

\institute{Institut f\"ur Astrophysik, Georg-August Universit\"at G\"ottingen, 37077 G\"ottingen, Germany
\and Max-Planck-Institut f\"ur Sonnensystemforschung, Justus-von-Liebig-Weg 3, 37077 G\"ottingen, Germany}

\date{Received <date> /
Accepted <date>}

\abstract {Local helioseismology has detected spatially extended converging surface flows into solar active regions. These play an important role in flux-transport models of the solar dynamo.}
{We aim to validate the existence of the inflows by deriving horizontal flow velocities around active regions with local correlation tracking of granulation.}
{We generate a six-year long-time series of full-disk maps of the horizontal velocity at the solar surface by tracking granules in continuum intensity images provided by the {\it Helioseismic and Magnetic Imager} (HMI) onboard the {\it Solar Dynamics Observatory} (SDO).}
{On average, active regions are surrounded by inflows extending up to $10^\circ$ from the center of the active region of magnitudes of 20--30~m/s, reaching locally up to 40~m/s, which is in agreement with results from local helioseismology. By computing an ensemble average consisting of 243 individual active regions, we show that the inflows are not azimuthally symmetric but converge predominantly towards the trailing polarity of the active region with respect to the longitudinally and temporally averaged flow field.}
{}

\maketitle


\section{Introduction}
Active regions are surrounded by spatially extended near-surface inflows~\citep{2001IAUS..203..189G,2002ApJ...570..855H,2004SoPh..224..217G,2004SoPh..220..371H,2004ApJ...613.1253H,2007ApJ...667..571K,2009ApJ...698.1749H,2011JPhCS.271a2007B} with typical flow velocities of $20 - 30$ m/s (velocities up to 50~m/s have been reported) that extend up to $10^\circ$ from the active region. Models suggest that these flows arise due to an enhanced cooling rate in the active regions~\citep{2003SoPh..213....1S,2008SoPh..251..241G,2008ApJ...684L.123H}. There are indications of outflows at greater depths~\citep[see e.g.,][and references therein]{2010ARA&A..48..289G}.

In flux-transport simulations, the inflows to active regions play an important role in the transport of magnetic flux over the course of the solar cycle. The inflows can affect the dispersal of active regions, since they counterbalance the outward diffusion of magnetic flux by convection~\citep{2006ESASP.624E..12D,2016A&A...586A..73M}. They are also a potential mechanism for modulating the strength of the solar cycle~\citep{2012A&A...548A..57C,2015ApJ...799..220S,2016A&A...586A..73M}.

So far, this inflow has only been studied using local helioseismology. Here we want to measure the inflows to active regions with local correlation tracking of solar granulation (LCT). Starting from continuum images provided by the {\it Helioseismic and Magnetic Imager}~\citep[HMI;][]{2012SoPh..275..229S}, we generate a time-series of full disk maps of the horizontal velocity at the solar surface by using the LCT code Fourier Local Correlation Tracking (FLCT)~\citep{2004ApJ...610.1148W,2008ASPC..383..373F}.

In order to demonstrate the viability of our technique, we measure a known flow of similar magnitude, the torsional oscillations~\citep{1980ApJ...239L..33H}, which have contemporaneous measurements by a different technique. In addition, we present results for individual active regions and for the average of 243 regions.

\section{Generating LCT full-disk flow maps}

\subsection{Running the FLCT code}
We compute maps of the horizontal velocity on the solar surface using the local correlation tracking code FLCT~\citep[Fourier Local Correlation Tracking;][]{2004ApJ...610.1148W,2008ASPC..383..373F}. FLCT provides the velocity by cross-correlating the granulation pattern between two consecutive intensity images. The FLCT code has been used for various applications, such as studying flows connected to solar activity~\citep[e.g.,][]{2009ApJ...705..821W,2013ApJ...766...39M,2013NatPh...9..489S} or systematic errors inherent to local correlation tracking~\citep{2016A&A...590A.130L}.

We run the FLCT code directly on full-disk images of the continuum intensity provided by HMI. The time-series stretches over six years of HMI data, from 24 April 2010 to 27 April 2016. In order to save computation time, we do not make use of the full cadence of HMI. We apply the FLCT code to pairs of continuum images with the images in the individual pairs separated by 45~s but the pairs separated by 30~min. This results in one flow map every 30~min. This setup slightly increases the noise in the flow maps that originates from the proper motion and evolution of the granules. However, the dominant part of the granulation noise is located at significantly smaller spatial scales than are relevant for this study~\citep[see also Fig. 3 in][]{2016A&A...587A...9L}. The FLCT code divides the images into several subimages, with a size given by the parameter $\sigma$. Here we use $\sigma = 6$~pixel.

Since we have not remapped the intensity images before running the FLCT code, the velocities provided by the code do not correspond to physical velocities on the Sun. Instead, they are defined on the grid of the CCD and have units of pixel per second with $v_m$ denoting the velocity in the horizontal direction (pointing from the left to the right) and $v_n$ denoting the velocity in the vertical direction (pointing from the bottom to the top). We convert the velocities from CCD to heliographic coordinates later in the analysis.

LCT generally underestimates the true flow velocities on the Sun~\citep[see e.g.,][]{2007SoPh..241...27S,2013A&A...555A.136V,2016A&A...587A...9L}. This effect depends on the spatial resolution of the data, hence, it varies across the disk. We correct for this effect by generating calibration data~\citep{2016A&A...587A...9L} both for $v_m$ and $v_n$. These calibration data consist of a time-series of continuum intensity images where we shift the individual images using Fourier interpolation in order to add a constant flow. The velocities estimated by the FLCT code depends linearly on the input velocities, with the slope ranging between $0.80$ and $0.89$.

The sensitivity of LCT also depends strongly on the magnetic field strength. For strong fields ($B \gtrsim 500$~G), the sensitivity of LCT drops significantly (down to $\sim 0.5$), since granulation is suppressed by the magnetic field. This reduction of the sensitivity cannot be easily corrected for because it depends strongly on disk position. This causes the background subtraction (see section below) to not work in regions with strong magnetic field. Hence, our LCT velocities are not reliable within and near sunspots.

\subsection{Subtracting the background signal}
\begin{figure}
\resizebox{\hsize}{!}{\includegraphics{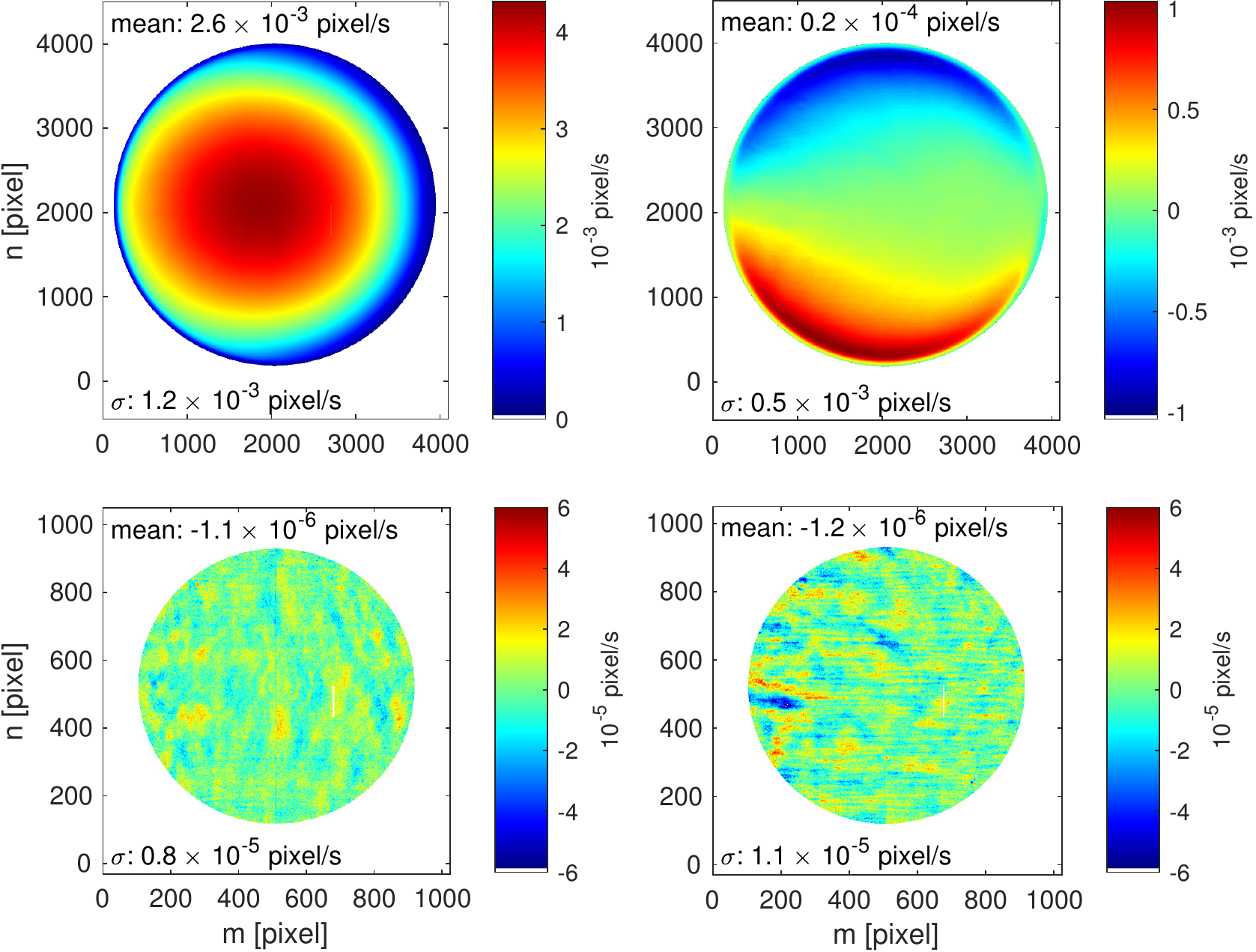}}
\caption{Mean flow maps generated by LCT without subtracting the background ({\it top}) and with background subtraction and $4\times 4$ subsampling ({\it bottom}). {\it Left:} flow maps for $v_m$, {\it right:} flow maps for $v_n$. The average consists of data between 24 April 2010 and 27 April 2016. The numbers in the {\it top left} of the individual subimages give the mean of the velocity, the numbers in the {\it bottom left} the standard deviation across the averaged image. At disk center, $10^{-5}$~pixel/s correspond to $\sim 3.50$~m/s and $10^{-3}$~pixel/s to $\sim 350$~m/s. The CCD of HMI has a bad column (top panels, $m=2727$), where the amplitude of the flow velocities in the $m$-direction are too low. We do not include these data in the analysis. In the {\it bottom} panels, we only show data within $60^\circ$ from disk center, we do not include data outside $60^\circ$ in the further analysis.}
\label{fig:mean_images}
\end{figure}
The LCT full disk flow maps are dominated by a large-scale background signal (see the mean images in the top panel of Fig.~\ref{fig:mean_images}), which needs to be removed in order to study the flows around active regions. This background consists of real solar flows (differential rotation and meridional flow) and a systematic error that looks like a flow converging towards disk center, also known as the shrinking-Sun effect~\citep{2004ESASP.559..556L,2016A&A...590A.130L}.

The background flow is time-dependent due to the spacecraft motion of SDO. In particular, the way that the flow maps are affected by solar rotation depends strongly on the heliographic latitude of disk center ($B_0$-angle). The position of a given solar latitude (with its specific rotation velocity) on the CCD changes with the $B_0$-angle. In addition, in the case of large values of the $B_0$-angle, differential rotation also affects the flow maps for $v_n$. The shrinking-Sun effect is also affected by the orbit of SDO. The $B_0$-angle exhibits predominantly an annual variation due to the Earth's orbit around the Sun. The orbit of SDO around the Earth causes additional temporal variations of the background flow on the time-scale of 24~h. Due to these temporal variations we cannot remove the background signal by subtracting a mean image from the flow maps.

The orbit of SDO is known, so in principle, it should be possible to model its influence on the LCT velocities. However, such an approach does not fully succeed in reproducing the 24~h variations \citep[compare with][who tried to model the influence of the orbit of SDO on the Doppler velocity]{2016ApJ...823..101S,2016SoPh..291.1887C}. In particular, the response of the shrinking-Sun effect to the orbit is only poorly understood, as this systematic effect is extremely sensitive to changes of the spatial resolution. Instead, we fit an empirical model to the background signal that smoothly varies in time and space. We model the background flow with Zernike polynomials $Z^m_n$~\citep{Zernike}, a set of orthogonal basis functions on the area of a circle (see Appendix~\ref{sect:Zernikes}). In our model, the coefficients of the individual Zernike polynomials vary smoothly in time in order to account for variations due to the orbit.

Removing the background signal using Zernike polynomials consists of several steps. First, we smooth the flow maps by convolving them with a Gaussian ($\sigma = 8$~pixel) followed by subsampling them $4\times 4$ in order to reduce the computation time. Then, we decompose each individual flow map both for $v_m$ and $v_n$ in Zernike polynomials. Since LCT becomes less reliable when used close to the limb, we fit the Zernike polynomials to the inner part of the disk (up to a heliocentric angle of $60^\circ$). We do not use data at heliocentric angles greater than $60^\circ$ in the further analysis. The center of the disk used for the fitting is the center of the solar disk determined from the limb, not the center of the CCD. Of course, the Zernike polynomials are not only sensitive to the background signal but also to the solar flows we are interested in. Hence, we do not subtract the Zernike polynomials from each flow map individually. Instead, our estimate of the background signal consists of a time-series for each Zernike-coefficient, stretching over all the data. We filter these data in Fourier space, selecting only the frequencies which we consider to represent the temporal evolution of the background signal (i.~e. signals with a period of 24~h or one year and the corresponding harmonics, see Appendix~\ref{sect:Zernikes_time} for more details). The signal at frequencies that are not affected by the orbit is not included in the time-series. After filtering the time-series of the Zernike polynomials, the mean and the standard deviation of the individual time-series decrease significantly with increasing radial degree $n$ of the Zernike polynomials. We use Zernike polynomials up to $n=7$, where for all $m$ both the mean and the standard deviation are $\leq 10^{-5}$~pixel/s (this corresponds to $\sim 3.5$~m/s at disk center). We then compute the background signal for each flow map from the filtered time-series of the Zernike-coefficients and subtract it. Fig.~\ref{fig:mean_images} shows the mean flow maps both for $v_m$ and $v_n$ with the background subtracted.

Our estimate of the background signal only includes temporal variations on time-scales that are affected by the orbit of SDO. The flows around active regions evolve on a different time-scale (days to weeks), meaning that they are not affected by the background subtraction. In principle, it would also be possible to mask the active regions before decomposing the flow maps into Zernike polynomials. However, in this case, the Zernike polynomials would no longer form an orthogonal basis. The leakage between the different polynomials would depend on what area on the disk is masked. This would be time-dependent since the active regions evolve and move across the disk due to solar rotation. In addition, masking a significant fraction of the disk would cause the Zernike polynomials to be more sensitive to convective noise, which increases the background noise in the power spectrum.

The mean image for $v_n$ with the background subtracted (bottom right part of Fig.~\ref{fig:mean_images}), exhibits a streakiness that is not present in the image for $v_m$. These features are caused by supergranulation. Since we have not tracked the data with the solar rotation when computing the mean images, individual supergranules propagate across the visible disk.

In the mean image for $v_m$ in Fig.~\ref{fig:mean_images}, a part of one of the columns ($m=2727$) has a lower velocity than the neighboring pixels. This is a bad column of the CCD. We do not include these data in the analysis.

\subsection{Remapping the LCT velocities}
In the next step, we track and remap the flow maps from the CCD coordinates to heliographic coordinates in order to generate Carrington maps of the flow velocities. The resulting flow maps have a spatial resolution of $0.4^\circ$. We also convert the flow velocities from the reference frame of the CCD (coordinate vectors $\hat{e}_m$ and $\hat{e}_n$) to spherical coordinates (coordinate vectors $\hat{e}_\phi$ and $\hat{e}_\theta$, see Appendix~\ref{sect:projection}). In addition, we generate Carrington magnetograms using the same tracking rate and the same spatial resolution as for the LCT flow maps.

\section{Flows derived using LCT}
\subsection{Carrington maps} \label{sect:Carrington}
\begin{figure*}
\centering
\includegraphics[width=17cm]{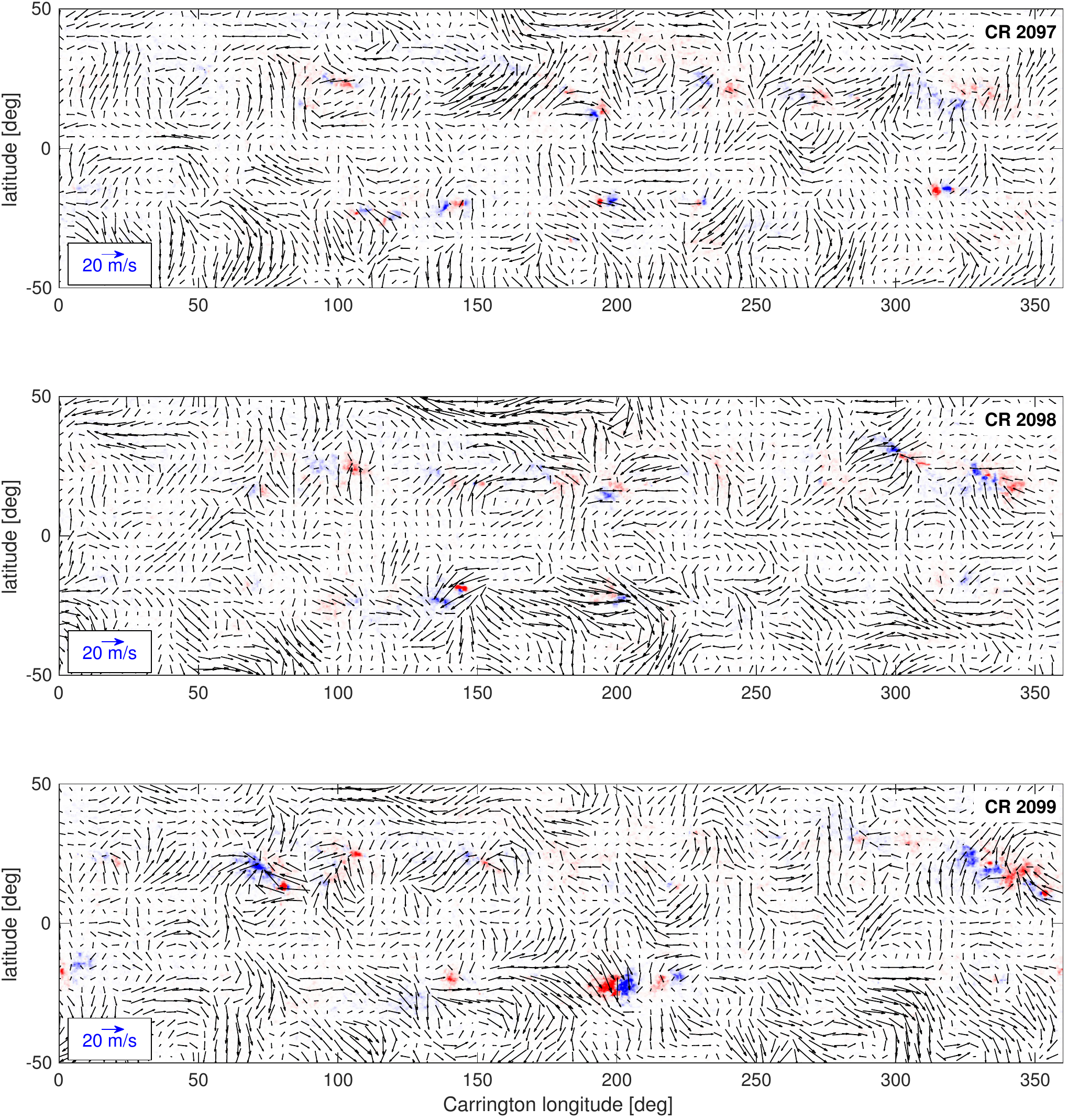}
\caption{Examples for Carrington maps of flows generated with LCT. The images show the line-of-sight magnetic field, the arrows show the LCT flow velocities with the background signal subtracted. The color scale saturates at $\pm 100$~G. The magnitude of the flow velocities is visualized in the lower left corner of the individual Carrington maps. We show three maps for the quiet Sun (CR~2097 from 20 May to 16 June 2010, CR~2098 from 16 June to 16 July 2010, and CR~2099 from 13 July to 9 August 2010). The rms of the flows is $\sim 8$~m/s.}
\label{fig:Carrington1}
\end{figure*}

\begin{figure*}
\centering
\includegraphics[width=17cm]{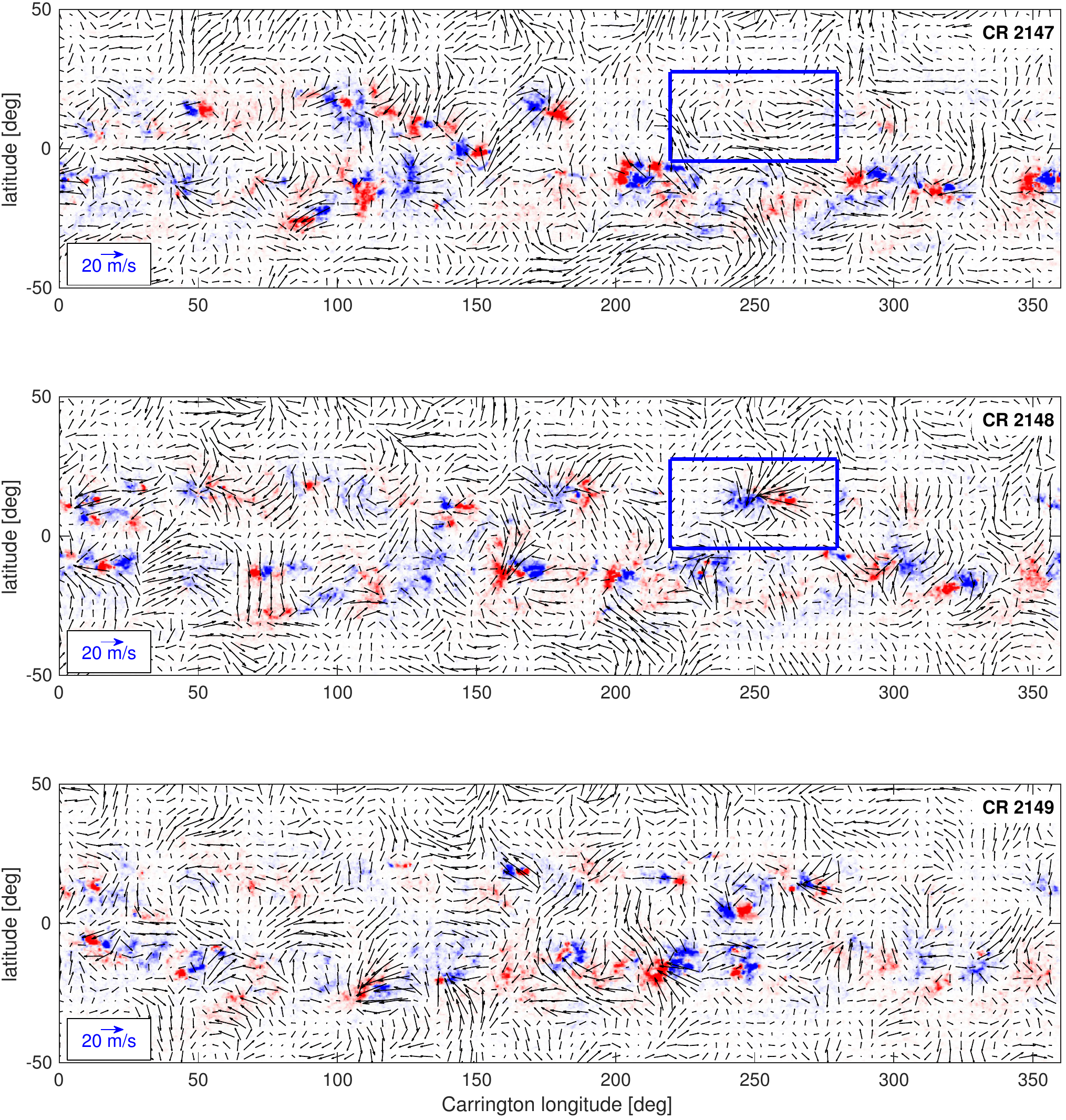}
\caption{Same as Fig.~\ref{fig:Carrington1} for three Carrington rotations near solar maximum, CR~2147 from 11 February to 11 March 2014, CR~2148 from 11 March to 7 April 2014 and CR~2149 from 7 April to 4 May 2014. The rms of the flows is $\sim 7$~m/s. The {\it blue boxes} outline an active region (AR~12007) that emerges between CR~2147 and CR~2148 on the far-side of the Sun and persists over the entire disk passage in CR~2148 without significant evolution. We study this active region in more detail in Figs.~\ref{fig:AR_evolution1} and~\ref{fig:AR_evolution2}.}
\label{fig:Carrington2}
\end{figure*}

Figs.~\ref{fig:Carrington1} and~\ref{fig:Carrington2} show examples of Carrington maps of the line-of-sight component of the magnetic field (averages of the magnetic field over the entire disk passage). We present results for three Carrington rotations at solar minimum (CR 2097, CR 2098, and CR 2099 in Fig.~\ref{fig:Carrington1}) with only a few active regions and three examples of Carrington maps at solar maximum (CR 2147, CR 2148, and CR 2149 in Fig.~\ref{fig:Carrington2}) with a high coverage by active regions. Since the velocities provided by LCT are dominated by velocities on small spatial scales (proper motion of granulation, supergranulation) we smooth the flow maps by convolving them with a Gaussian ($\sigma = 4^\circ$). In addition, we subtract for each latitude the longitudinally and temporally averaged flow velocity (both for $v_\phi$ and $v_\theta$).
The Carrington maps exhibit a complex flow pattern, both in the quiet Sun and near active regions. In the quiet Sun, there are several examples of cyclonic inflows and anticyclonic outflows. A prominent feature is visible in the Carrington rotations 2097-2099 at $\sim45^\circ$ latitude and $\sim150^\circ$ longitude. In~\citet{2013Sci...342.1217H}, this flow cell was also detected and identified as a giant cell. Most active regions in Figs.~\ref{fig:Carrington1} and~\ref{fig:Carrington2} are surrounded by converging inflows with typical flow velocities of 20-30~m/s. The inflows are far from being azimuthally symmetric, extending locally up to $10^\circ$ from the active region and reaching flow velocities of up to 40~m/s.

\subsection{Torsional oscillations}
\begin{figure*}
\centering
\includegraphics[width=17cm]{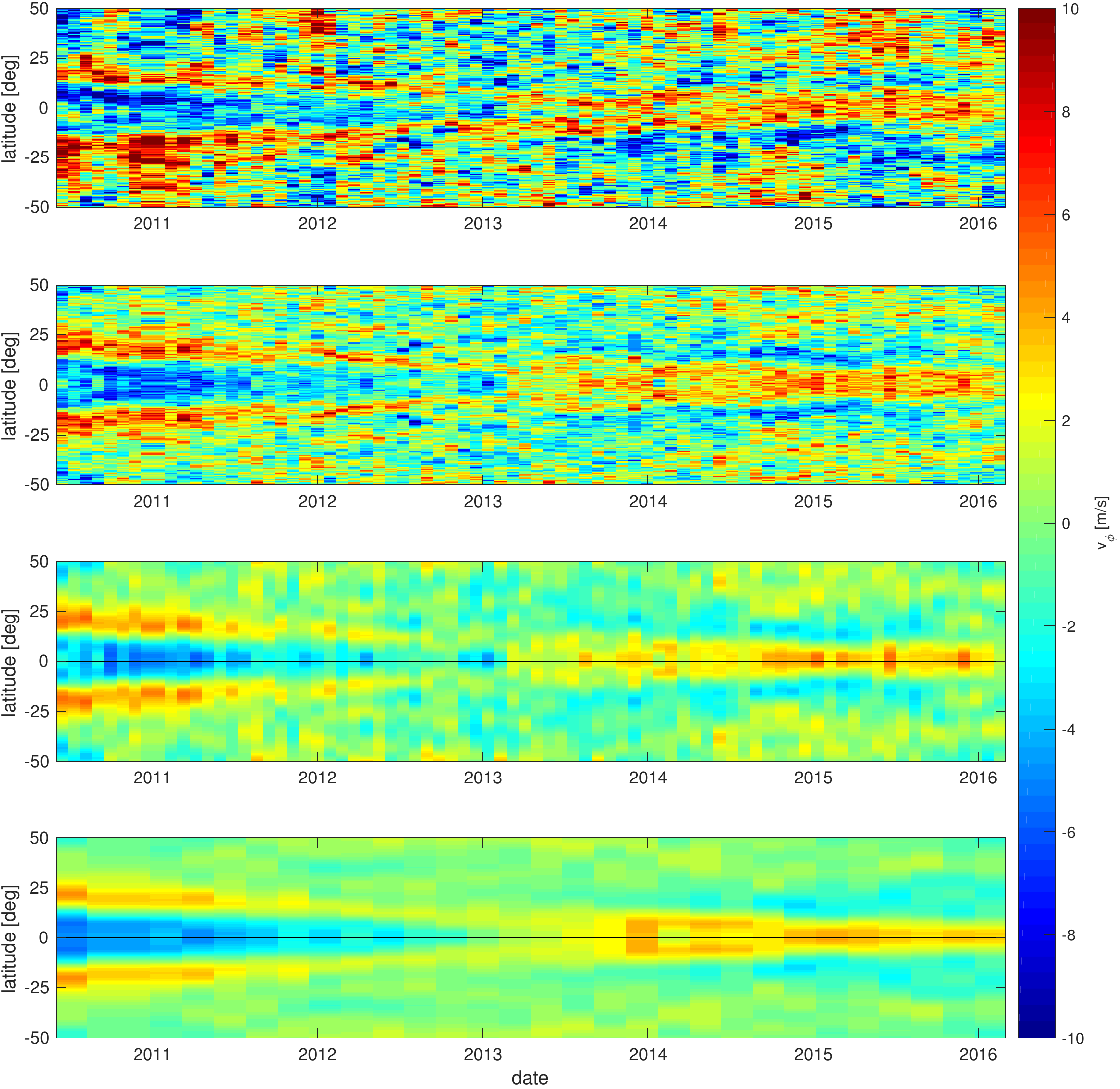}
\caption{Comparison of the torsional oscillations derived using the LCT flow maps ($v_\phi$) presented in this study and from global helioseismology with the f-mode~\citep[provided by][]{f-mode}.  From {\it top} down: nominal LCT $v_\phi$, LCT $v_\phi$ symmetrized around the equator, LCT $v_\phi$ symmetrized and convolved with the latitudinal component of the averaging kernels of the f-mode study. The {\it bottom} panel shows the results from the f-mode data. We have subtracted the mean rotation rates for all data sets, respectively.}
\label{fig:torsional_oscillations}
\end{figure*}

\begin{figure}
\resizebox{\hsize}{!}{\includegraphics{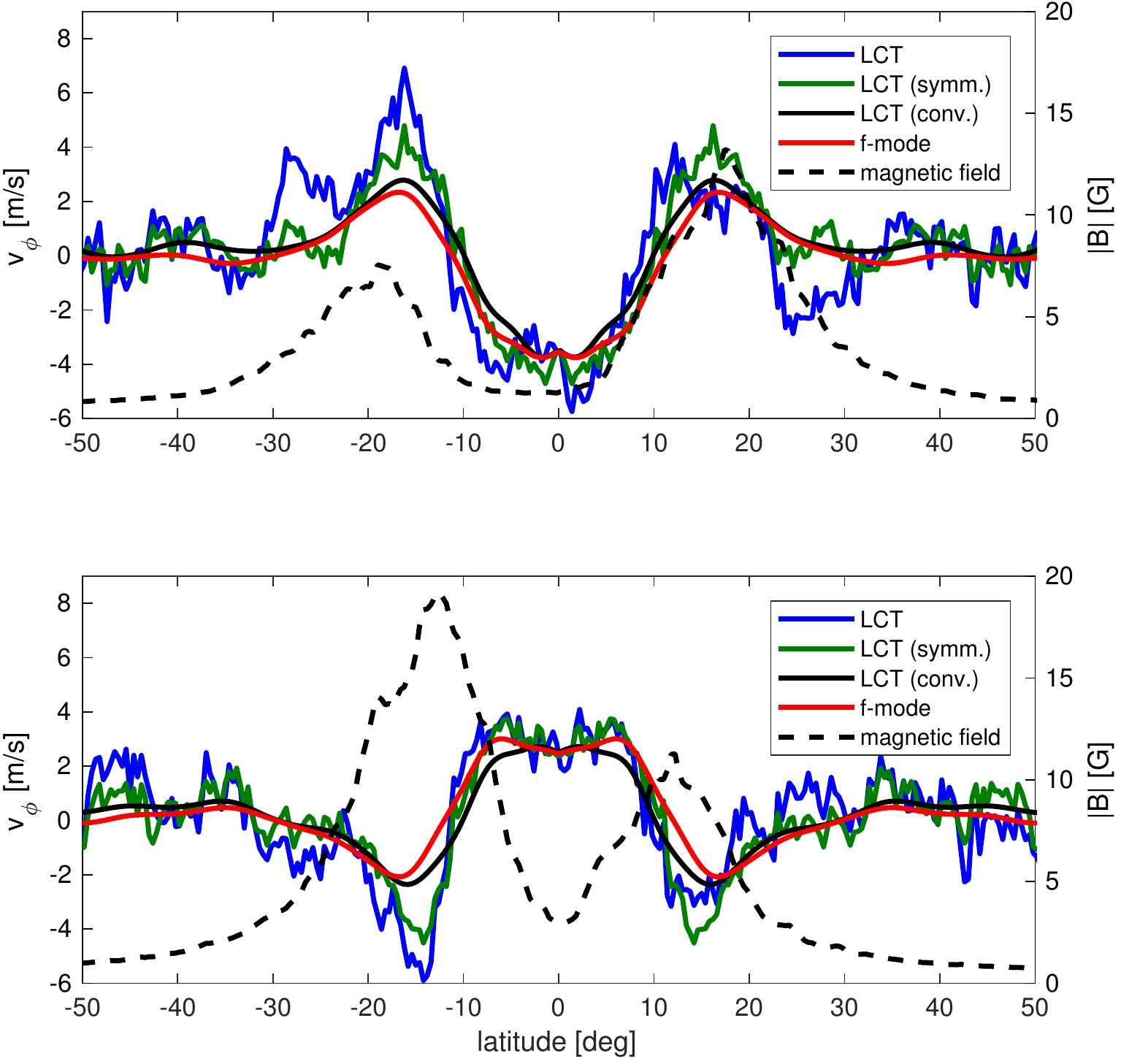}}
\caption{Annual averages of the torsional oscillation data ($v_\phi$) shown in Fig.~\ref{fig:torsional_oscillations}. The {\it top} panel shows data for 2011, the {\it bottom} panel shows data for 2014. The {\it blue} curve shows the nominal LCT data, the {\it green} curve shows the LCT data after symmetrizing in latitude, the {\it black} curve shows the LCT data after symmetrizing them and convolving them with the averaging kernels of the f-mode data, the {\it red} curve shows the results from global helioseismology with the f-mode, and the {\it dashed black} curve shows the averaged unsigned magnetic field.}
\label{fig:torsional_oscillations_cut}
\end{figure}

In order to test the quality of the LCT flow maps, we compare the torsional oscillations seen in our data and the flows seen by global mode helioseismology. These are inversions of the f-mode frequency splittings using HMI full-disk data \citep{f-mode} and were processed as described in \citet{1999ApJ...523L.181S}. Fig.~\ref{fig:torsional_oscillations} and~\ref{fig:torsional_oscillations_cut} show the flows in the $\phi$-direction after subtracting the mean flow at each latitude from both data sets. The LCT data exhibits signal at smaller spatial scales than the f-mode data and also contains the component of the torsional oscillations that is antisymmetric with respect to the equator (global helioseismology is only sensitive to the symmetric component of the flow). However, after symmetrizing the LCT data around the equator and smoothing them with the averaging kernels of the global helioseismology data, the results derived from the two methods are in good agreement and the annual averages shown in Fig.~\ref{fig:torsional_oscillations_cut} have a comparable rms over time ($\sigma = 0.3$~m/s).


\subsection{Significance of the inflows: case study of AR~12007}
\begin{figure*}
\centering
\includegraphics[width=17cm]{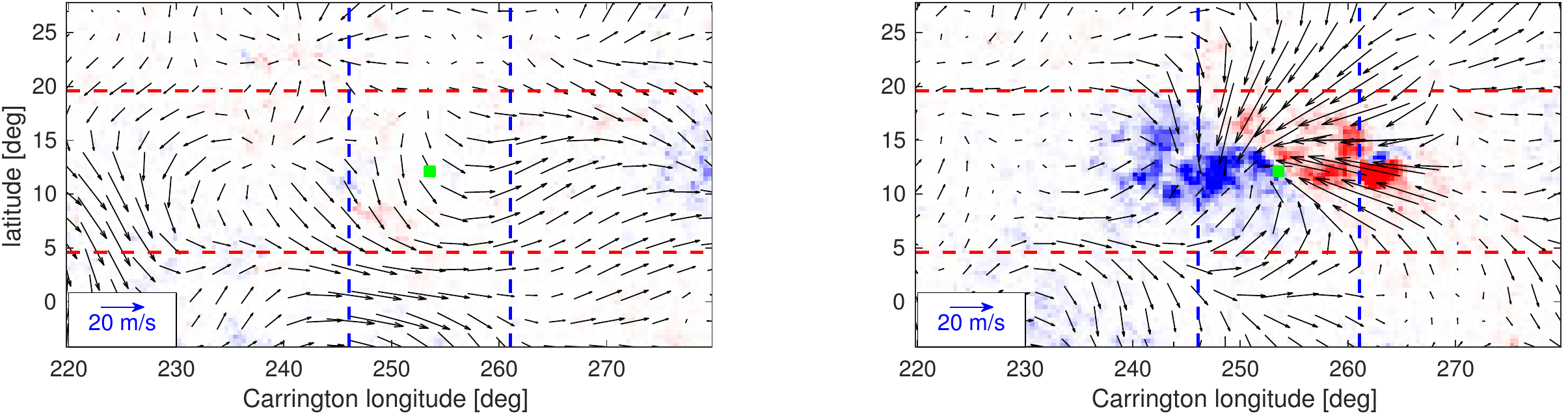}
\caption{Comparison of the flows around a selected active region (AR~12007, the active region highlighted by the blue boxes in Fig.~\ref{fig:Carrington2}) one Carrington rotation before emergence (CR~2147, {\it left panel}) and one Carrington rotation after emergence (CR~2148, {\it right panel}). It emerges between CR~2147 and CR~2148 on the far-side of the Sun and persists over the entire disk passage in CR~2148 without significant evolution. The images show the line-of-sight magnetic field, the arrows show the LCT flow velocities. The color map saturates at $\pm 100$~G, the magnitude of the flow velocities is visualized in the {\it lower left} corner of the individual subimages. The {\it dashed lines} indicate strips in latitude ({\it red horizontal lines}) and in longitude ({\it blue vertical lines}) around the center-of-gravity of the unsigned magnetic field (indicated by the {\it green squares}). In Fig.~\ref{fig:AR_evolution2}, we show averages of the longitudinal and latitudinal velocities along these strips.}
\label{fig:AR_evolution1}
\end{figure*}

\begin{figure}
\centering
\resizebox{\hsize}{!}{\includegraphics[width=17cm]{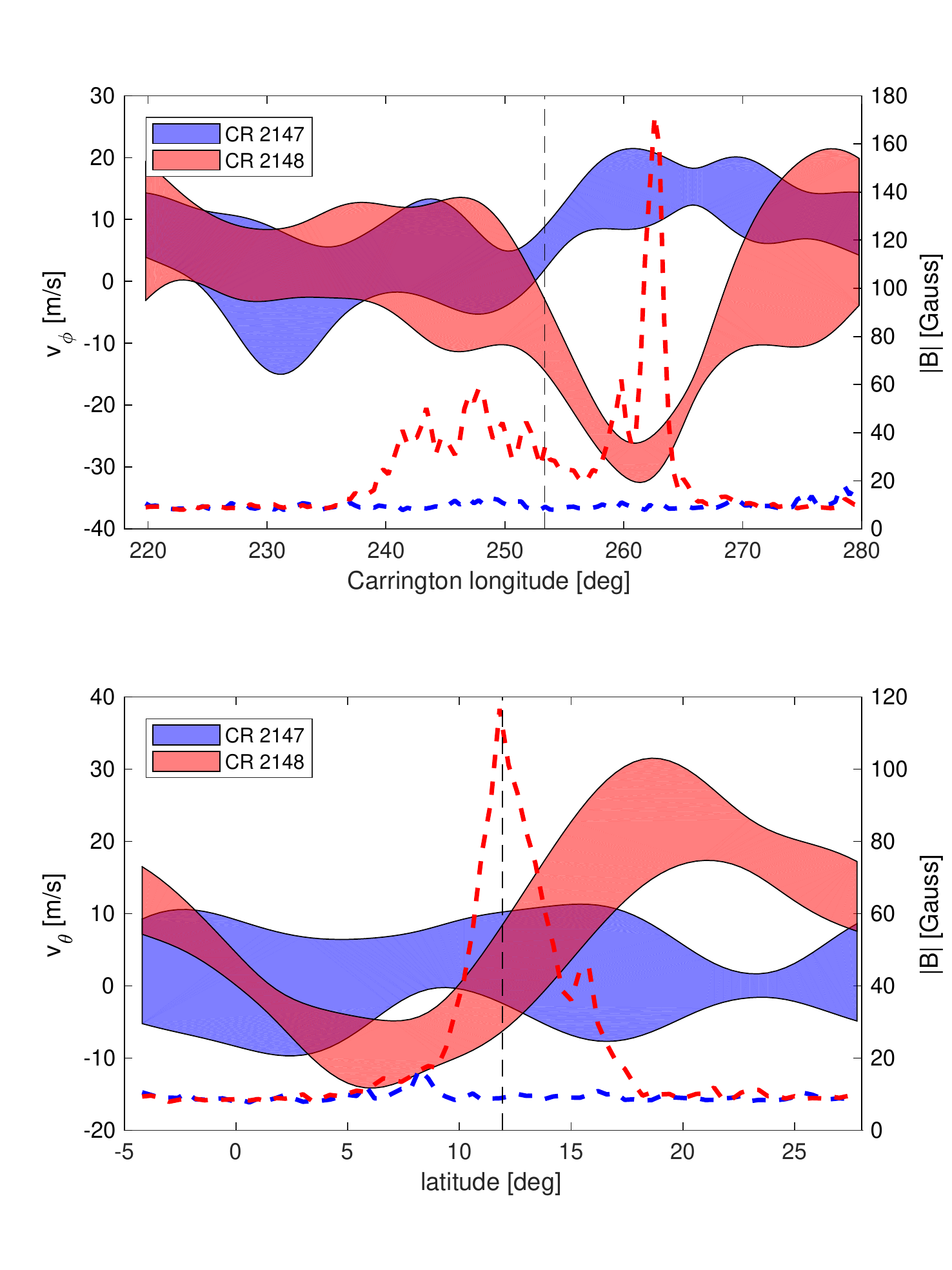}}
\caption{Averages of the flow velocities of the active region AR~12007 shown in Fig.~\ref{fig:AR_evolution1}. We show averages of $v_\phi$ along a strip in longitude ({\it top}) and of $v_\theta$ along a strip in latitude ({\it bottom}). Both strips are $15^\circ$ wide and centered around the center-of-gravity of the unsigned magnetic field of the active region (see {\it dashed lines} in Fig.~\ref{fig:AR_evolution1}). The {\it shaded areas} show the flow velocities with the width corresponding to the standard-deviation of the velocity over five consecutive 24~h averages of the flow velocities for the individual Carrington rotations. The {\it dashed lines} show the averaged unsigned magnetic field. {\it Blue}: CR~2147, {\it red}: CR~2148. The {\it dashed vertical lines} indicate the position of the center-of-gravity of the unsigned magnetic field (derived from the data of CR~2148).}
\label{fig:AR_evolution2}
\end{figure}

The flows around active regions are a superposition of the inflows and convective noise (mostly supergranulation). Even after smoothing the flow maps with the Gaussian ($\sigma = 4^\circ$), the rms level ($\sim 8$~m/s) is comparable to the magnitude of the inflows. This requires a test whether the inflow signal is real or just caused by noise. We test this for the active region AR~12007 (highlighted by the blue boxes in Fig.~\ref{fig:Carrington2}). This active region emerges on the far-side of the Sun between CR~2147 and CR~2148 and persists over the entire disk passage in CR~2148 without significant evolution. Fig.~\ref{fig:AR_evolution1} shows the flows around this active region one Carrington rotation after emergence (CR~2148) and the flows in the same region on the quiet Sun one Carrington rotation before the emergence (CR~2147). After emergence, there is an inflow to the active region, while there is no sign of a converging flow in the Carrington map before emergence. However, the magnitude of the flow velocities is similar in both Carrington maps. In order to test if the inflow signal is real and is not just noise, we show in Fig.~\ref{fig:AR_evolution2} the daily variation of $v_\phi$ and $v_\theta$ over the course of one Carrington rotation after averaging them along strips in latitude and longitude with a width of $15^\circ$ which are centered around the center-of-gravity of the unsigned magnetic field of the active region (between the dashed lines in Fig.~\ref{fig:AR_evolution1}). In the quiet Sun (CR~2147), the flow velocities are dominated by short-lived noise (predominantly supergranulation). After the emergence of the active region, the noise level is not changed but an inflow towards the active region has developed. The temporal variation of the flows is consistent with the noise level in the quiet Sun, indicating that the inflow does not undergo substantial temporal evolution and is indeed a real signal.


\subsection{Flows around the averaged active region}
\begin{figure}
\centering
\resizebox{\hsize}{!}{\includegraphics[width=17cm]{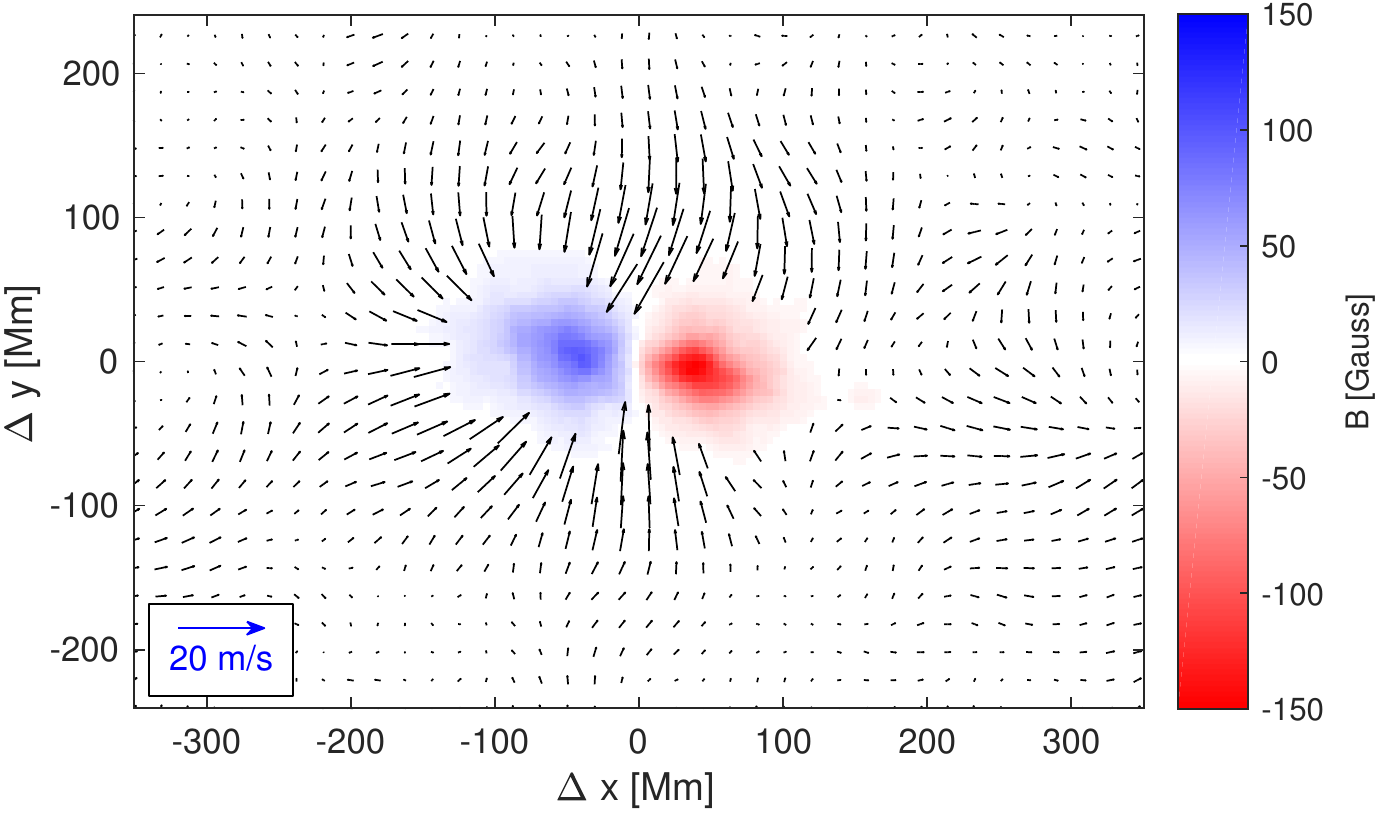}}
\caption{Flow velocities and magnetic field for an average of 243 individual active regions with strong magnetic flux. For details of the averaging procedure, see text. The noise of the averaged flow velocities is smaller than one m/s.}
\label{fig:AR_mean}
\end{figure}

As discussed in the previous section, the inflows to individual active regions are affected by supergranulation noise. This means that studying individual active regions is not sufficient for making general statements about the inflows. We address this issue by computing the flow field around an averaged active region. We automatically identify active regions with high magnetic flux and then add up the flow field around the selected active regions. This process consists of several steps.
We start by convolving Carrington maps of the unsigned magnetic field with a Gaussian ($\sigma = 2^\circ$). In these maps, we then identify all regions, where the unsigned magnetic field exceeds 30~Gauss as active regions. In total, we detect 971 active regions. Since we are only interested in active regions with a high total magnetic flux, we then integrate the unsigned magnetic field over the area of the individual active regions. In the following, we consider only those active regions, where the total flux is within the top 25\% (larger than $5.9\times 10^{21}$~Mx). This results in 243 active regions. We then average the magnetic field and the flow field for these selected active regions by aligning the center-of-gravity of the unsigned magnetic field (active regions on the southern hemisphere are flipped in latitude and the sign of the magnetic field is switched). We do not consider the flow velocities within the individual active regions in order to avoid contamination of the flow maps by the magnetic field. Also, when averaging, we do not take into account that the individual active regions have different sizes, tilt angles, or shapes.
Fig.~\ref{fig:AR_mean} shows the magnetic field and the flow field for the resulting averaged active region. The averaged active region clearly exhibits an inflow that extends up to 200~Mm from the center of the active region and reaches magnitudes of $\sim 15$~m/s. This is lower than for individual active regions and is caused by averaging over active regions with different sizes and shapes. The inflow around the averaged active region is far from being azimuthally symmetric. The inflow converges predominantly to the trailing polarity. Also, most of the inflow occurs in the $\theta$-direction, with higher latitudes contributing more than lower latitudes. The shape of this mean inflow depends on the tracking rate and on the subtraction of the background flows. As stated in Sect.~\ref{sect:Carrington}, we have subtracted the longitudinally and temporally averaged flow velocities from each latitude before averaging the flow velocities.

The magnetic field also exhibits an asymmetry. In the leading polarity, the magnetic field reaches up to 146 Gauss, while it is only 97 Gauss in the trailing polarity. Morphological asymmetries of active regions have been reported before~\citep[see e.g., review by][]{Fan2004}.

\section{Discussion}
We generated a six-year long time-series of full-disk maps of horizontal flows at the solar surface generated using LCT. This includes a new approach for removing systematic errors. When systematic errors exhibit a strong time dependence, e. g., due to variations of the $B_0$-angle, the commonly used approach of subtracting a mean image does not work very well. This method requires averaging the data over a certain amount of time in order to reduce the noise level to a satisfactory level. During this time, however, the systematic error might change. Our approach of describing the spatial dependency with Zernike polynomials (or any other suitable set of functions) and then filtering the resulting time-series of the model parameters in Fourier space reduces this problem. It takes into account the time-evolution of the systematic errors while not being affected by noise.

Our dataset covers a long time-series and a large fraction of the solar surface. In addition, LCT allows us to resolve a broad range of spatial scales, from supergranulation up to large-scale flow patterns. This means that these data can be used for studying a broad range of science topics. As an example, we use the data to study the inflows to active regions. We can, for the first time, confirm the existence of the inflows with a method that is independent from local helioseismology. In addition, by averaging over a large sample of active regions, we reveal an asymmetry of the inflows. The flows converge predominantly to the trailing polarity of the active region and are predominantly in the $\theta$-direction (with respect to the longitudinally and temporally averaged flow velocities). This effect has not been taken into account by the simple parametrized models used for flux-transport dynamo simulations~\citep{2006ESASP.624E..12D}. Studying the statistics of the inflows over a large number of active regions with different parameters (magnetic flux, latitude, time since emergence, etc.) could also help us test different physical models for the origin of the inflows.

\begin{acknowledgements}
We acknowledge support from Deutsche Forschungsgemeinschaft (DFG) through SFB 963/1 "Astrophysical Flow Instabilities and Turbulence" (Project A1). Support was also provided by European Union FP7 projects SPACEINN and SOLARNET. The German Data Center for SDO, funded by the German Aerospace Center (DLR), provided the IT infrastructure for this project. L.G. acknowledges support from the Center for Space Science, NYU Abu Dhabi Institute, UAE.
\end{acknowledgements}

\bibliographystyle{aa} 
\bibliography{literature} 

\appendix

\section{Definition of the Zernike polynomials}\label{sect:Zernikes}
Zernike polynomials $Z_n^m (\rho,\varphi)$~\citep{Zernike} form a set of orthogonal functions on a disk of radius one. Here, $n$ (radial degree, $\geq 0$) and $m$ (azimuthal degree) are integers with $n \geq |m|$. The variables $\rho$ and $\varphi$ are the radial distance and the azimuthal angle respectively. The Zernike polynomials are defined in the following way:
$$Z_n^m (\rho,\varphi)=\begin{cases} \sqrt{2n+2} \ R_n^m (\rho) \cos{(m\varphi)}& \textmd{for }  m>0 \\ \sqrt{n+1} \ R_n^m (\rho)& \textmd{for } m=0 \\ \sqrt{2n+2} \ R_n^m (\rho) \sin{(m\varphi)}& \textmd{for } m<0.\end{cases}$$
The radial component $R_n^m (\rho)$ is defined as:
$$R_n^m(\rho) = \sum_{k=0}^{\frac{n-m}{2}} \frac{(-1)^k (n-k)!}{k! \left (\frac{n+m}{2}-k \right )! \left ( \frac{n-m}{2}-k \right )!} \rho^{n-2k}.$$
The Zernike polynomials are only defined if $n-m$ is even. 

We choose to normalize the Zernike polynomials as follows:
$$\int_0^1 \int _0^{2\pi} Z_n^m (\rho,\varphi) Z_{n'}^{m'} (\rho,\varphi) \rho d\rho d\phi = \pi \delta_{nn'}\delta_{mm'}.$$

\section{Time-evolution of the Zernike coefficients}\label{sect:Zernikes_time}
\begin{figure}
\resizebox{\hsize}{!}{\includegraphics{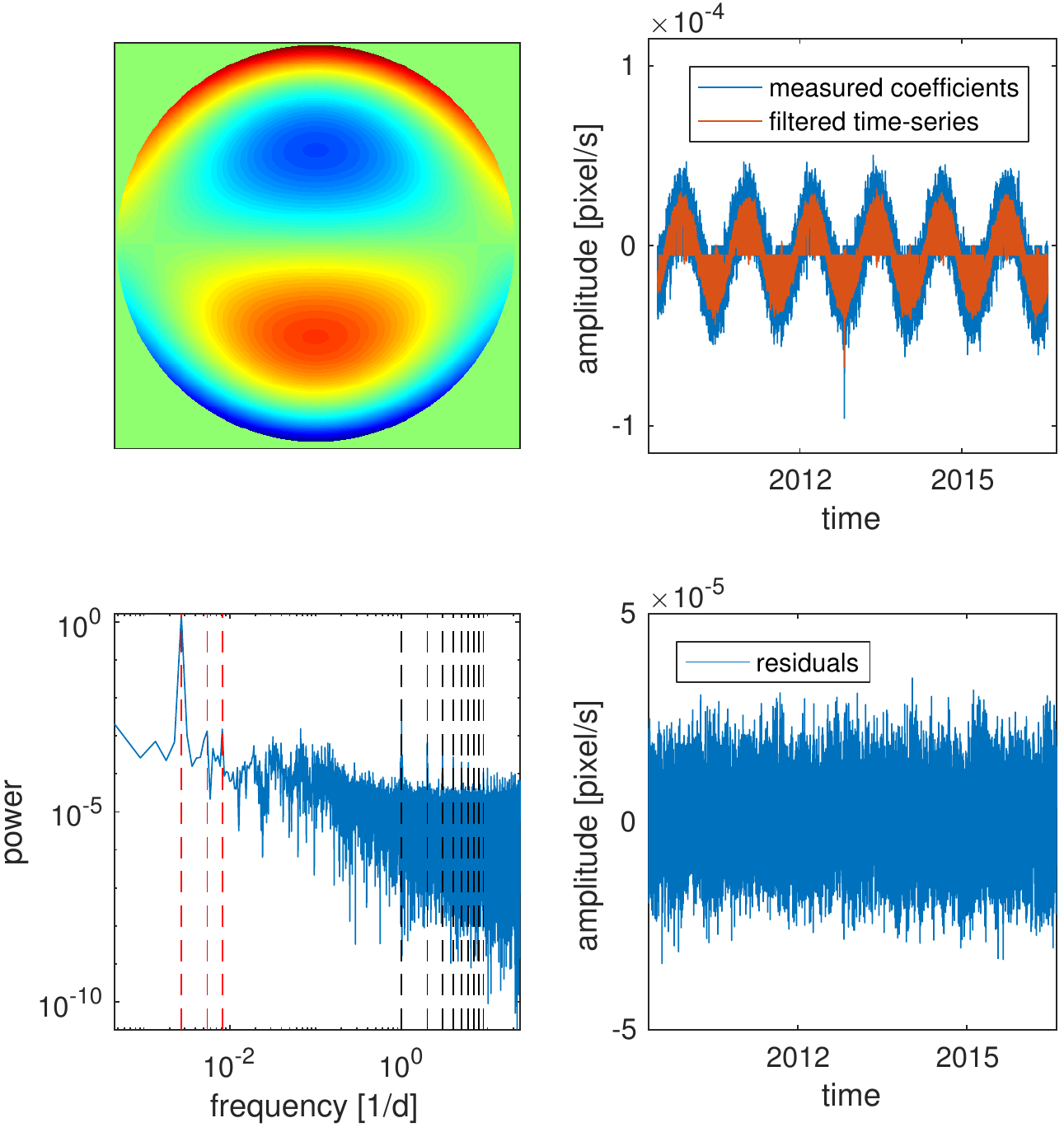}}
\caption{Time-evolution of the coefficient for the Zernike polynomial with $n=3$ and $m=-1$ for $v_m$. The {\it top left} panel visualizes this Zernike polynomial, the {\it top right} panel shows the time-series of the amplitude of the coefficient, both before ({\it blue curve}) and after ({\it red curve}) filtering in time. The {\it bottom left} panel shows the power spectrum of the unfiltered time-series. The dashed vertical lines indicate examples of frequencies that are affected by the orbit, by the orbit of the Earth around the Sun ({\it red lines}) or by the orbit of SDO around the Earth ({\it black lines}). For more details about the filtering, see text. The residuals after subtracting the filtered from the unfiltered time-series are shown in the {\it bottom right} panel.}
\label{fig:Zernikes}
\end{figure}

The top right part of Fig.~\ref{fig:Zernikes} shows the time-evolution of the Zernike coefficients for the Zernike polynomial with $n=3$ and $m=-1$ (shown in the top left part in Fig.~\ref{fig:Zernikes}) for $v_m$. The time-series is dominated by a yearly oscillation due to variations of the $B_0$-angle. In addition, the orbit of SDO causes a series of peaks in the power spectrum at a period of 24~h and its harmonics. Since the orbit of SDO around the Earth is modulated by the orbit of the Earth around the Sun, each of these harmonics is a multiplet consisting of several individual peaks. These peaks are separated by a frequency of $\sim 2.8\times 10^{-3}$~1/day, which corresponds to a period of one year. Apart from that, the power is almost constant, suggesting that it originates from real solar flows or white noise. The time-series for the individual Zernike coefficients do not all look the same. Not all Zernike coefficients exhibit a yearly period and which harmonics of the 24~h period can be seen is also not always the same.

Here, we account for the influence of the orbit using a purely empirical approach. For each Zernike-coefficient, we generate a time-series consisting of the signal measured at the frequencies which we consider to be affected by the orbit. First, we remove the extreme outliers from the time-series. In some cases, the HMI data are corrupted (mostly due to eclipses), which affects the flow maps. If the coefficient for $n=0$ and $m=0$ (the mean of the flow map) for $v_m$ or for $v_n$ deviates by more than $5\sigma$ from the mean, we assume this time step to be corrupted and replace all Zernike coefficients for that time step by their respective mean values. We do not include these data in our analysis. We perform the outlier correction two times consecutively. In the next step, we compute the Fourier transform of the time-series and filter for only those frequencies that are affected by the orbit of SDO plus the mean of the time-series. These are the daily and yearly variations plus the corresponding harmonics. In case of the yearly variations, the main frequency ($2.8\times 10^{-3}$~1/day) plus the first 16 harmonics (up to a frequency of $4.4\times 10^{-3}$~1/day) exhibit significant power and are taken into account. In case of the daily variations, we use the main frequency (1/day) and all 23 harmonics that can be resolved with the cadence of our data (30~min, frequency of the last harmonic: 24/day). Since the daily variations and its harmonics are not isolated peaks but multiplets with the individual peaks being separated by a frequency of 1/year, we extract the signal in a range in frequency around the central frequency ($\pm 7.5\times 10^{-3}$~1/day) in order to cover the entire multiplet. Taking account only the signal at the frequencies selected in this manner, we finally generate a time-series for each Zernike-polynomial (see Fig.~\ref{fig:Zernikes} for an example).

This approach works for most Zernike polynomials, but not for all of them. The time-series for the polynomial with $n=0$ and $m=0$ (the mean of the velocity) both for $v_m$ and for $v_n$ are affected by the orbit on all temporal scales. In these two cases, we do not select specific frequencies but use the full frequency spectrum for the further analysis. Issues arise also for the polynomials with $n=1$ and $m=-1$ for $v_m$ and  $n=1$ and $m=1$ for $v_n$. The power spectra of the time-series for these polynomials exhibit a significantly higher background power than the other Zernike polynomials. The time-series for these two Zernike coefficients are highly correlated, suggesting that this background noise originates from rotating the images. It is known that the $p$-angle of the HMI data exhibits oscillations whose origin is not fully understood~\citep{2016A&A...587A...9L}. We account for this by performing a principal component analysis of the two time-series. We assume that the correlated component originates from the $p$-angle oscillations, while the uncorrelated part represents real solar flows.

\section{Converting the eigenvectors from CCD to spherical coordinates}\label{sect:projection}
The velocities derived using LCT are defined on a 2D Cartesian coordinate system that is aligned with the plane of the CCD. The $m$-axis points to the right, and the $n$-axis points upwards. Tracking and remapping these velocities to heliographic coordinates results in $v_m(\lambda,\phi)$ and $v_n(\lambda,\phi)$, with $\lambda$ and $\phi$ being the heliographic latitude and longitude, respectively. The velocities are now given as a function of the heliographic coordinates, but their coordinate vectors, $\hat{e}_m$ and $\hat{e}_n$, are still defined on the 2D Cartesian coordinate system of the CCD. In addition, the velocities are still given in units of pixel/s and need to be converted to m/s. In order to transform the coordinate vectors from CCD to spherical coordinates, we need to make the assumption that we only measure velocities on the surface of the Sun (i.~e. $r=R_\odot$) and that they are purely horizontal (i.~e. $v_r = 0$).

In the first step, we take into account the finite distance of SDO to the Sun and convert the velocities from pixel/s to m/s. We define a new coordinate system centered on the Sun given by the basis vectors $\hat{x'}$, $\hat{y'}$, and $\hat{z'}$, with the $x'$-axis pointing towards the observer, the $\hat{y'}$-axis pointing to the right ($\hat{m}$-direction), and the $z'$-axis pointing upwards ($\hat{n}$-direction). The coordinates in the system are related to the CCD coordinates by
\begin{align}
m &= y' \frac{1}{s(D - x')}, \\
n &= z' \frac{1}{s(D - x')}.
\end{align}
Here, $s$ is the plate scale of HMI (in rad/pixel) and $D$ is the distance from SDO to the center of the Sun. In order to convert the velocities, we take the partial derivative of these equations with respect to time and solve for $v_y'$ and $v_z'$. Since we assume our velocities to be purely horizontal, we can express $v_x'$ as a function of $v_y'$ and $v_z'$. This leads to the following equations:
\begin{align}
v_{y'} &= \frac{s (D-x') \left [ J_{11} (D-x') v_m + J_{13} (v_n y' - v_m z')\right ]}{J_{11} (D-x') - (J_{12} y' + J_{13} z')}, \label{eq:1} \\
v_{z'} &= \frac{s (D-x') \left [ J_{11} (D-x') v_n - J_{12} (v_n y' - v_m z')\right ]}{J_{11} (D-x') - (J_{12} y' + J_{13} z')}. \label{eq:2}
\end{align}
Here, the elements of the matrix $J$ originate from expressing $v_x'$ as a function of $v_y'$ and $v_z'$ (in order to enforce $v_r = 0$) and are given by:
\begin{align}
J_{11} &= \sin(\theta) \cos(\phi) \cos(B_0) + \cos(\theta) \sin(B_0),\\
J_{12} &= \sin(\theta) \sin(\phi),\\
J_{13} &= -\sin(\theta) \cos(\phi) \sin(B_0) + \cos(\theta) \cos(B_0).
\end{align}
In the next step, we rotate the coordinate system given by the basis vectors $\hat{x'}$, $\hat{y'}$, and $\hat{z'}$ around the $y'$-axis by an angle $-B_0$. In the resulting coordinate system with unit vectors $\hat{x}$, $\hat{y}$, and $\hat{z}$, the $z$-axis is aligned with the rotation axis of the Sun. Then, we transform from the Cartesian coordinate system, $\hat{x}$, $\hat{y}$, and $\hat{z}$, to spherical coordinates $\hat{r}$, $\hat{\phi}$ (direction of increasing heliocentric longitude), and $\hat{\theta}$ (direction of increasing co-latitude). Again, we use the assumption that that the velocities on the Sun are purely horizontal (i.~e. $v_r = 0$). This results in the following equations for $v_\phi$ and $v_\theta$:
\begin{align}
v_\phi &= \frac{1}{A} (a_{11}v_y' + a_{12} v_z'), \label{eq:3} \\
v_\theta &= \frac{1}{A} (a_{21}v_y' + a_{22} v_z'). \label{eq:4}
\end{align}
The elements of the matrix $a$ and the factor $A$ are given by:
\begin{align}
a_{11} &= \cos{\theta} \cos{\phi} \sin{B_0} + \sin{\theta} \cos{B_0},\\
a_{12} &= \cos{\theta} \sin{\phi},\\
a_{21} &= \sin{\phi}\sin{B_0},\\
a_{22} &= -\cos{\phi},\\
A &= \sin{\theta} \cos{\phi} \cos{B_0} + \cos{\theta} \sin{B_0}.
\end{align}
Inserting equations \ref{eq:1} and \ref{eq:2} in \ref{eq:3} and \ref{eq:4} gives the final result
\begin{align}
v_\phi &= c_{11} v_m + c_{12} v_n,\\
v_\theta &= c_{21} v_m + c_{22} v_n.
\end{align}
The coefficients are given by
\begin{align}
c_{11} &= C \cos{\phi} ( D \cos{\theta} \sin{B_0}-R_\odot) + C D \sin{\theta}\cos{B_0},\\
c_{12} &= C(D\cos{\theta}\sin{\phi}-R_\odot\sin{\phi}\sin{B_0}),\\
c_{21} &= C(D\sin{\phi}\sin{B_0}-R_\odot \cos{\theta}\sin{\phi}),\\
c_{22} &= C\cos{\phi} (R_\odot \cos{\theta} \sin{B_0}-D)+CR_\odot \sin{\theta}\cos{B_0},\\
C &= \frac{s (D-R_\odot \cos{\theta} \sin{B_0}-R_\odot \sin{\theta}\cos{\phi}\cos{B_0})}{D\cos{\theta}\sin{B_0}+D\sin{\theta}\cos{\phi}\cos{B_0}-R_\odot}.
\end{align}

\end{document}